\documentstyle[12pt]{article}
\def\be {\begin{equation}}
\def\ee {\end{equation}}
\def\ba {\begin{eqnarray}}
\def\ea {\end{eqnarray}}

\def\b  {\beta}

\def\p  {\pi}

\def\r  {\rho}

\def\la {\label}
\def\le {\left}
\def\ri {\right}
\def\pa {\partial}
\def\f {\frac}

\def\bi {\begin{itemize}}
\def\ei {\end{itemize}}
\begin{document}
\def\bea{\begin{eqnarray}}
\def\eea{\end{eqnarray}}
\title{\bf {Logarithmic correction to the Cardy-Verlinde formula in Topological
 Reissner-Nordstr\"om de Sitter Space }}
 \author{M.R. Setare  \footnote{E-mail: rezakord@ipm.ir}
  \\{Physics Dept. Inst. for Studies in Theo. Physics and
Mathematics(IPM)}\\
{P. O. Box 19395-5531, Tehran, IRAN }\\
 and \\ {Department of Science, Physics group, Kordestan
University, Sanandeg,  Iran }}

%\date{\small{\today}}
\maketitle
\begin{abstract}
In this paper we compute leading order correction due to small
statistical fluctuations around equilibrium, to the Cardy-Verlinde
entropy formula (which is supposed to be an entropy formula of
conformal field theory in any dimension) of a Topological
Reissner-Nordstrom black hole in de Sitter space.
 \end{abstract}
% \begin{document}
\newpage
% \vspace*{10mm}

 \section{Introduction}
 It is commonly believed that any valid theory of quantum gravity
 must necessary incorporate the Bekenestein-Hawking definition of
 black hole entropy \cite{bek,haw} into its conceptual framework.
 However, the microscopic origin of this entropy remains an enigma
 for two reasons. First of all although the various counting
 methods have pointed to the expected semi-classical result, there
 is still a lack of recognition as to what degrees of freedom are
 truly being counted. This ambiguity can be attributed to most of
 these methods being based on dualities with simpler theories,
 thus obscuring the physical interpretation from the perspective of
 the black hole in question. Secondly, the vast and varied number
 of successful counting techniques only serve to cloud up an
 already fuzzy picture.\\
 de Sitter/Conformal Field Theory correspondence (dS/CFT) may hold
 the key to its microscopical interpretation. Naively, we would expect dS/CFT
 correspondence to proceed along the lines of Anti-de Sitter /Conformal Field Theory
  (AdS/CFT) correspondence because de Sitter spacetime  can be
  obtained from anti-de Sitter spacetime by analytically continuing
  the cosmological constant to imaginary values. However, local
  and global properties of dS spacetime lead to unexpected
  obstructions. Unlike AdS, the boundary of dS spacetime is
  spacelike and its dual CFT is Euclidean. Moreover, dS spacetime
  does not admit a global timelike Kiling vector. The time
  dependence of the spacetime metric precludes a consistent
  definition of energy and the use of Cardy formula to compute dS
  entropy. Finally dS/CFT duality leads to boundary operators with
  complex conformal weights, i.e. to a non-unitary CFT. In spite
  of these difficulties, some progress towards a consistent definition of dS/CFT
   correspondence has been achieved ~\cite{AS}-\cite{cad2}\\
   There has been much recent interest in calculating the quantum
corrections to $S_{BH}$ (the Bekenestein-Hawking entropy)
\cite{maj1x}-\cite{medved}. The leading-order correction is
proportional to $\ln{S_{BH}}$. There are, {\it two} distinct and
separable sources for this logarithmic correction
 \cite{gg2x,maj3x} (see also recent paper by Gour and Medved \cite{ medved}).
  Firstly, there should be a correction
 to the number of microstates that is a quantum correction to the
 microcanonical entropy, secondly, as any black hole will typically exchange heat or
 matter with its surrounding, there should also be a correction due to thermal
 fluctuations in the horizon area.
 In this paper we consider Topological Reissner-Nordstrom black hole in de Sitter space
 in arbitrary dimension. In section $2$ we calculate the
 corresponding thermodynamical quantities for both cosmological
 and black hole horizon. In section $3$ at first we review the
 calculation of the logarithmic correction due to the thermal
 fluctuation to the entropy, then we obtain the corresponding
 correction to the Cardy-Verlind formula. Last section contain a
 summary of paper.

\section{Thermodynamical quantities of TRNdS black hole}
The topological Reissner-Nordstr\"om dS black hole solution in
$(n+2)$-dimensions has the following form
\begin{eqnarray}
&& ds^2 = -f(r) dt^2 +f(r)^{-1}dr^2 +r^2 \gamma_{ij}dx^{i}dx^{j}, \nonumber \\
&&~~~~~~ f(r)=k -\frac{\omega_n M}{r^{n-1}} +\frac{n \omega_n^2
Q^2}{8(n-1) r^{2n-2}}
     -\frac{r^2}{l^2},
\end{eqnarray}
where
\begin{equation}
\omega_n=\frac{16\pi G}{n\mbox {Vol}(\Sigma)},\hspace{1cm}\phi
=-\frac{n}{4(n-1)}\frac{\omega_n Q}{r^{n-1}},
\end{equation}
where $Q$ is the electric/magnetic charge of Maxwell field, $M$ is
assumed to be a positive constant, $l$ is the curvature radius of
de Sitter space, $\gamma_{ij}dx^idx^j$ denotes the line element of
an $n$-dimensional hypersurface $\Sigma_k$ with the constant
curvature $n(n-1)k$ and its volume $V(\Sigma_k)$. $\Sigma_k$ is
given by spherical ($k=1$), flat ($k=0$), hyperbolic  $(k=-1)$,
$\phi$ is the elctrostatic potential related to the charge $Q$.
When $k=1$, the metric Eq.(1) is just the Reissner-Nordstr\"om-de
Sitter solution. For general $M$ and $Q$, the equation $f(r)=0$
may have four real roots. Three of them are real, the largest on
is the cosmological horizon $r_{c}$, the smallest is the inner
(Cauchy) horizon of black hole, the middle one is the outer
horizon $r_{+}$ of the black hole. And the fourth is negative and
has no physical meaning. The case $M=Q=0$ reduces
to the de Sitter space with a cosmological horizon $r_{c}=l$.\\
When $k=0$ or $k=-1$, there is only one positive real root of
$f(r)$, and this locates the position of cosmological horizon
$r_{c}$.\\
In the case of $k=0$, $\gamma_{ij}dx^{i}dx^{j}$ is an
$n-$dimensional Ricci flat hypersurface, when $M=Q=0$ the solution
Eq.(1) goes to pure de Sitter space
\begin{equation}
ds^{2}=\frac{r^{2}}{l^{2}}dt^{2}-\frac{l^{2}}{r^{2}}dr^{2}+r^{2}dx_{n}^{2},
\end{equation}
in which $r$ becomes a timelike coordinate.\\
When $Q=0$, and $M\rightarrow -M$ the metric Eq.(1)is the TdS
(Topological de Sitter) solution \cite{med}, which have
a cosmological horizon and a naked singularity.\\
In the BBM prescription \cite{BBM}, the gravitational mass,
subtracted the anomalous Casimir energy, of the TRNdS solution is
\begin{equation}
\label{3eq3} E^{c}=-M =-\frac{r_c^{n-1}}{\omega_n} \left (k
-\frac{r_c^2}{l^2} +
    \frac{n\omega_n^2 Q^2}{8(n-1)r_c^{2n-2}}\right).
\end{equation}
 The Hawking temperature $T_{TRNdS}^{c}$
and entropy $S_{TRNdS}^{c}$ associated with the cosmological
horizon are
\begin{eqnarray}
 && T_{TRNdS}^{c}=\frac{-f'(r_{c})}{4\pi} =\frac{1}{4\pi r_c} \left(-(n-1)k +(n+1)\frac{r_c^2}{l^2}
    +\frac{n\omega_n^2 Q^2}{8 r_c^{2n-2}}\right), \nonumber \\
&& S
_{TRNdS}^{c}=\frac{r_c^n\mbox{Vol}(\Sigma)}{4G},\label{cosentr}
\end{eqnarray}
where $V_{c}=r_{c}^{n}Vol(\Sigma)$ is area of the cosmological
horizon. The AD mass of TRNdS solution can be expressed in terms
of black hole horizon radius $r_+$ and charge $Q$,
\begin{equation}
 E^{b} =M =\frac{r_+^{n-1}}{\omega_n} \left
(1-\frac{r_+^2}{l^2} +
   \frac{n\omega_n^2 Q^2}{8(n-1)r_+^{2n-2}}\right).
\end{equation}
The black hole horizon Hawking temperature $T_{TRNdS}^{b}$ and
entropy $S_{TRNdS}^{b}$ are given by
\begin{eqnarray}
 && T_{TRNdS}^{b}=\frac{f'(r_{+})}{4\pi} =\frac{1}{4\pi r_+} \left((n-1) -(n+1)\frac{r_+^2}{l^2}
   -\frac{n\omega_n^2 Q^2}{8 r_+^{2n-2}}\right), \nonumber \\
&& S_{TRNdS}^{b}
=\frac{r_+^n\mbox{Vol}(\Sigma)}{4G},\label{blacentr}
\end{eqnarray}
where $r=r_{+}$ is black hole horizon and
$V_{+}=r_{+}^{n}Vol(\Sigma)$ is area of it in $(n+2)-$dimensional
asymptotically dS space. Therefore, there are two kinds of
temperatures corresponding to two horizon, the system is not
thermodynamically stable. However, the system should be adiabatic
since one can define the temperature in the vicinity of each
horizon.\\
 Making use of the fact that the metric for the
boundary CFT can be determined only up to a conformal factor, we
rescale the boundary metric for the CFT to be of the following
form
\begin{equation}
ds^2_{CFT}=\lim_{r\to\infty}\left[{l^2\over r^2}ds^2_{n+2}\right]
=-dt^2+l^2\gamma_{ij}dx^idx^j.
\end{equation}
Then the thermodynamic relations between the boundary CFT and the
bulk TRNdS are given by
\begin{eqnarray}
E_{CFT}&=&M{l\over r},\ \ \ \ \ \ \ \ \ \ \Phi_{CFT}=\Phi{l\over
r},
 \cr T_{CFT}&=&{T_{TRNdS}}{l\over r},\ \ \ \ \ \ \
 S_{CFT}=S_{TRNdS},
\end{eqnarray}
 \\
 The specific heat of the black hole is given by
 \begin{equation}
 C^{c,b}=\frac{dE^{c,b}}{dT}=\frac{4\pi r_{c,+}^{2}(8 k (1-n)l^{2}r_{c,+}^{n-2}+8 (n+1)r_{c,+}^{n}+
 n\omega_{n}^{2}l^{2}r_{c,+}^{-n}Q^{2}) }
 { \omega_{n}( 8l^{2}(n-1)k+8
 r_{c,+}^{2}(n+1)+(1-2n)l^{2}\omega_{n}^{2}r_{c,+}^{2-2n}Q^{2})}.
 \end{equation}
 As one can see the above specific heat is positive in the case $k=-1,
 k=0$, for $k=1$, $C^{c,b}$ is positive only with following condition
 \begin{equation}
8 (n+1)r_{c,+}^{n}+n\omega_{n}^{2}l^{2}r_{c,+}^{-n}Q^{2} > 8 k
(1-n)l^{2}r_{c,+}^{n-2}
 \end{equation}
\section{Logarithmic correction to the entropy and Cardy-Verlinde formula}
At first we review the calculation of the logarithmic correction
to the entropy (see the paper by Das et al. \cite{das2x}), the
partition function in the canonical ensemble is given by \be Z(\b)
%= \sum_i \r(E_i) e^{-\b E_i}
= \int_0^\infty \r(E) e^{-\b E} dE~~, \label{part1} \ee
where $T=1/\b$ is the temperature in units of the Boltzmann
constant $k_B$. The density of states can be obtained from
(\ref{part1}) by doing an inverse Laplace transform (keeping $E$
fixed)
\be \r(E) = \f{1}{2\p i} \int _{c-i\infty}^{c+i\infty} Z(\b) e^{\b
E} d\b = \f{1}{2\p i} \int_{c-i\infty}^{c+i\infty} e^{S(\b)}d\b~~,
\la{density1} \ee
where
\be S(\b) = \ln Z(\b) + \b E \label{basic} \ee is the {\it exact}
entropy as a function of temperature, not just its value at
equilibrium. The complex integral can be performed by the method
of steepest descent around the saddle point $\b_0 (= 1/T_0)$, such
that $S'_0 := (\pa S(\b)/\pa \b)_{\b=\b_0}=0$. $T_0$ is the
equilibrium temperature, such that the usual equilibrium relation
$E=-(\pa \ln Z(\b)/\pa \b)_{\b=\b_0}$ is obeyed. Expanding $S(\b)$
around $\b=\b_0$, we get
\be S = S_0 + \f{1}{2} (\b - \b_0) ^2 S_0'' + \cdots ~~,
\label{ent1} \ee
Substituting (\ref{ent1}) in (\ref{density1}) :

\be \r(E) = \f{e^{S_0}}{\sqrt{2\p S''_0}} ~~. \la{corr0} \ee

Note that the density of states $\r(E)$ and $S_0''$ have
dimensions of inverse energy and energy squared respectively.
Henceforth, we set the Boltzmann constant $k_B$ to unity. The
logarithm of the density of states $\r(E)$ is then the
microcanonical entropy \be {\cal S} = \ln \r(E) = S_0 - \f{1}{2}
\ln S_0'' + ~\mbox{(higher order terms)}. \la{corr1} \ee
 Now, we will estimate $S_0''$, without
assuming any specific form of $S(\b)$. From Eq.(\ref{basic}), it
follows that \be S''(\b)={1\over Z}({\pa^2 Z(\b)\over {\pa
\b^2}})- {1\over Z^2}({\pa Z\over \pa \b})^2~. \ee This means that
$S_0''$ is nothing but the fluctuation squared of energy from the
equilibrium, i.e.,
\be S_0''= <E^2>-<E>^2~, \ee
where, by the definition of $\b_0$, $E=<E>= - (\pa \ln Z/\pa
\b)_{\b=\b_0}$. It immediately follows that \be S_0''= T^2 C
\label{esci} \ee
where $C \equiv (\partial E/ \partial T)_{T_0}$ is the
dimensionless specific heat. Substituting for $S_0''$ from
(\ref{esci}) in (\ref{corr1}), we get:
\be {\cal S} = \ln \r = S_0 - \f{1}{2} \ln \le( C~T^2 \ri) +
\cdots \la{corr3} \ee This equation can only be directly applied
if the specific heat is non negative. Therefore the entropy always
has the logarithmic correction due to thermal fluctuation. To get
entropy as a logarithm of dimensionless quantity, we can multiply
the density of state with ${\cal {E}}$ which has the dimension of
energy \cite{mukx}, then the correction to the entropy becomes
\begin{equation}
{\cal S}=S_0+{\rm ln }\frac{{\cal {E}}} {\sqrt{C T^2}}+\ldots
\label{entro1}
\end{equation}
we can set the scale $\cal{E}$ to be the temperature $T$ of the
system, this is because we have temperature as the only available
scale in canonical ensamble. Therefore we have
\begin{equation}
\label{entro} {\cal S}=S_0-\frac{1}{2}{\rm ln }{C}+\ldots
\end{equation}
When $r_{c,+}^{2}\gg l^{2}$, $C\simeq n S_0$, in this case we have
\begin{equation}
\label{entro1} {\cal S}=S_0-\frac{1}{2}{\rm ln }{S_0}+\ldots
\end{equation}
\\
 It is now
possible to drive the corresponding correction to Cardy-Verlinde
formula. The Casimir energy $E_c$, defined as
\begin{equation}\label{caseq}
 E_{C}^{c,b}=(n+1) E^{c,b}-nT^{c,b}S^{c,b}-n\phi^{c,b} Q,
 \end{equation}
 in this case, is found to be
 \begin{equation}
 E_{C}^{c,b} =\frac{-2n k r_{c,+}^{n-1}Vol(\Sigma)}{16\pi
G},
\end{equation}
which is valid for both cosmological and black hole horizon. One
can see that the entropy Eqs. (\ref{cosentr},\ref{blacentr}) of
the cosmological and black hole horizon can be written as
\begin{equation}
 S^{c,b}=\frac{2\pi l}{n}\sqrt{|\frac{E_{C}^{c,b}}{k}|(2(E^{c,b}-E_{q}^{c,b})-E_{C}^{c,b})},
\end{equation}
where
\begin{equation}
E_{q}^{c,b} = \frac{1}{2}\phi^{c,b} Q
=-\frac{n}{8(n-1)}\frac{\omega_n Q^2}{r_{c,+}^{n-1}}.
\end{equation}
 For the present discussion, the total entropy is assumed to be of
 the form Eq.(\ref{entro1}), where the uncorrected entropy, $S_{0}$
 correspondence to that associated in Eqs. (\ref{cosentr},\ref{blacentr}). It then follows by
 employing Eqs.(4, 5,6,7) that the Casimir energy Eq.(\ref{caseq}) can
 be expressed in term of the uncorrected entropy. (Following
 expressions are valid for both cosmological and black hole horizon, then for simplicity
 we omit the subscript $c$ and $b$ )
 \begin{equation}
  E_{C} =\frac{-2n r_{c,+}^{n-1}\mbox{Vol}(\Sigma)}{16\pi
G}+\frac{nT}{2}Ln S_{0},
\end{equation}
Then we obtain
\begin{equation}\label{sapro}
\frac{2\pi l}{n}\sqrt{|\frac{E_{C}}{k}|(2(E-E_{q})-E_{C})}\simeq
S_{0}-( \frac{r_{c,+}^{2}}{kl^{2}+1})\frac{\pi l^{2}T}{2r_{c,+}}Ln
S_0
\end{equation}
In the limit where the correction is small, the coefficient of the
logarithmic term on the right-hand side of Eq.(\ref{sapro}) can be
expressed in terms of the energy and Casimir energies
\begin{equation}
\frac{E-E_q-E_c/2 }{E_{c}/2}=\frac{-r_{c,+}^{2}}{kl^{2}}
\end{equation}
Using the following equation
\begin{equation}
(\frac{r_{c,+}^{2}}{kl^{2}}+1)\frac{\pi l^{2}}{2r_{c,+}}
\frac{((n+1)E-E_c-2nE_q)\omega_{n}}{4\pi
r_{c,+}^{n}}=(1-\frac{E_{c}/2}{E-E_q-E_{c}/2
})\frac{2nE_q+E_c-(n+1)E}{4E_c},
\end{equation}
we may conclude, therefore that in the limit where the logarithmic
corrections are sub-dominant, Eq.(\ref{sapro}) can be rewritten to
express the entropy in terms of the energy, energy of
electromagnetic and Casimir energy.
 \bea
 S_0&=&\frac{2\pi
l}{n}\sqrt{|\frac{E_c}{k}|(2(E-E_q)-E_c)}\\
 &&+
\frac{(2nE_q+E_c-(n+1)E)}{4E_c}\frac{(E-E_q-E_c)}
{E-E_q-E_{c}/2}Ln( \frac{2\pi
l}{n}\sqrt{|\frac{E_c}{k}|(2(E-E_q)-E_c)} )\nonumber \eea and
consequently, the total entropy Eq.(\ref{entro1}) to first order
in the logarithmic term, is given by \bea \label{corent}
 S &\simeq&\frac{2\pi
l}{n}\sqrt{|\frac{E_c}{k}|(2(E-E_q)-E_c)}+ \frac{E_{q}[(3n+1)E-2n
E_{q}+(1-2n)E_c]+E[nE_c-(n+1)E]}{4E_c(E-E_q-E_{c}/2)} \nonumber\\
&& Ln \left (\frac{2\pi l}{n}\sqrt{|\frac{E_c}{k}|(2(E-E_q)-E_c)}
\right) \eea Therefore taking into account thermal fluctuations
defines yhe logarithmic corrections to both cosmological and
black hole entropies. As a result the Cardy-Verlinde formula
receive logarithmic corrections in our interest  TRNdS black hole
background in any dimension, in the way similar to the
Cardy-Verlinde formula for the SAdS and SdS black holes in
5-dimension \cite{{od},{od1}}

  \section{Conclusion}
 For a large class of black hole, the Bekenstein-Hawking entropy
 formula receives additive logarithmic corrections due to thermal
 fluctuations. On the basis of general thermodynamic arguments, Das et al
 \cite{das2x} deduced that the black hole entropy can be expressed as the
 Eq.(\ref{corr3}). In this paper we have analyzed this correction of the entropy of TRNdS
 black hole in any dimension in the light of dS/CFT. We have obtain the logarithmic
 correction to both cosmological and black hole entropies. Then using the form of the
 logarithmic correction Eq.(\ref{entro1}) we
 have drived the corresponding correction to the Cardy-Verlinde formula which relates the
 entropy of a certain CFT to its total energy and Casimir energy in arbitrary dimension.

  \vspace{3mm}

\section*{Acknowledgement }
I would like to thank Saurya Das for useful discussions and
suggestions.


\vspace{3mm}








\begin{thebibliography}{99}
\bibitem{bek} J.D. Bekenstein, Lett. Nuovo. Cim. {\bf 4}, 737 (1972);
Phys. Rev. {\bf D7}, 2333 (1973); Phys. Rev. {\bf D9}, 3292
(1974).
\bibitem{haw} S.W. Hawking, Comm. Math. Phys. {\bf 25}, 152 (1972);
 J.M. Bardeen, B. Carter and S.W. Hawking, Comm. Math. Phys.
{\bf 31}, 161 (1973).
\bibitem{AS} A. Strominger, JHEP {\bf 0110}, 034, (2001);  M. Spradlin, A. Strominger,
 A. Volovich, hep-th/0110007.
\bibitem{dS}
C.M. Hull,  JHEP {\bf 9807}, 021, (1998).
\bibitem{Mazu}P.~O.~Mazur and E.~Mottola,
Phys.\ Rev.\ D {\bf 64}, 104022 (2001). I.~Antoniadis, P.~O.~Mazur
and E.~Mottola, astro-ph/9705200.
\bibitem{Bala}V.~Balasubramanian, P.~Horava and D.~Minic,
JHEP {\bf 0105}, 043, (2001).
\bibitem{mu1}Mu-In Park, Nucl. Phys. {\bf B544},377, (1999).
\bibitem{Li}M.~Li, JHEP {\bf 0204}, 005, (2002).
\bibitem{Noji2}S.~Nojiri and S.~D.~Odintsov,
Phys.\ Lett.\ B {\bf 519}, 145, (2001); JHEP {\bf 0112},
033,(2001); S.~Nojiri, S.~D.~Odintsov and S.~Ogushi, Phys. Rev.
{\bf D65}, 023521 (2002); hep/th0205187.
\bibitem{Klem3}D.~ D. Klemm, A. C. Petkou, G. Siopsis, Nucl. Phys. {\bf B601}, 380, (2001);
Klemm, Nucl. Phys. {\bf B625}, 295, (2002); S.~Cacciatori and
 D.~Klemm, Class. Quant. Grav. 19, 579, (2002).
\bibitem{Gao}Y.~h.~Gao,
hep-th/0107067.
\bibitem{Bros}J.~Bros, H.~Epstein and U.~Moschella, Phys. Rev. {\bf D65},
084012,(2002).
\bibitem{Haly}E.~Halyo,
hep-th/0107169.
\bibitem{set}M. R. Setare, Mod. Phys. Lett. {\bf A17}, 2089, (2002).
\bibitem{set1} M. R. Setare and R. Mansouri, Int. J. Mod. Phys. {\bf A18}, 4443, (2003).
\bibitem{set2}M. R. Setare, M. B. Altaie,  Eur. Phys. J.  {\bf C30},
 273, (2003).
\bibitem{set3}M . R. Setare and E. C. Vagenas, Phys. Rev.{\bf D68},
064014, (2003).
\bibitem{Toll}A.~J.~Tolley and N.~Turok,
hep-th/0108119.
\bibitem{Shir}T.~Shiromizu, D.~Ida and T.~Torii, JHEP {\bf 0111}, 010, (2001).
\bibitem{McIn}B.~McInnes, Nucl. Phys. {\bf B627}, 311, (2002).
\bibitem{Strom1}A.~Strominger, JHEP {\bf 0111}, 049, (2001).
\bibitem{BBM}V.~Balasubramanian, J.~de Boer and D.~Minic, Phys. Rev. {\bf D65},
123508,(2002).
\bibitem{myung1}Y.~S.~Myung, Mod. Phys. Lett. {\bf A16}, 2353, (2001).
\bibitem{Carn}B.~G.~Carneiro da Cunha, Phys. Rev. {\bf D65}, 104025,
(2002).
\bibitem{CMZ}R.~G.~Cai, Y.~S.~Myung and Y.~Z.~Zhang, Phys. Rev. {\bf D65}, 084019,
(2002).
\bibitem{Dani}U.~H.~Danielsson, JHEP {\bf 0203}, 020, (2002).
\bibitem{gez1}A. M. Ghezelbash, R. B. Mann, JHEP {\bf 0201}, 005
(2002); A. M. Ghezelbash, D. Ida, R. B. Mann, T. Shiromizu, Phys.
Lett.{\bf B535}, 315, (2002).
\bibitem{deh} M. H. Dehghani, Phys. Rev. {\bf D66},044006 (2002);
ibid. {\bf D65}, 104030 (2002); ibid.  {\bf D65}, 104003, (2002).
\bibitem{Ogus}S.~Ogushi, Mod. Phys. Lett. {\bf A17}, 51, (2002).
\bibitem{med}A.J.M. Medved, Class. Quant. Grav. {\bf 19}, 2883,
(2002).
\bibitem{med1}A.J.M. Medved, Class. Quant. Grav. {\bf 19}, 919, (2002).
\bibitem{cad1}M. Cadoni, P. Carta, S. Mignemi, Nucl.Phys. {\bf B632}, 383, (2002).
\bibitem{cad2}M. Cadoni, P. Carta, M. Cavaglia', S. Mignemi, Phys. Rev. {\bf D66},  065008,
(2002).
\bibitem{maj1x} R.K.  Kaul and P. Majumdar, Phys. Rev. Lett. {\bf 84}, 5255
(2000).
\bibitem{makx} J. Makela and P. Repo, gr-qc/9812075.
\bibitem{carx} S. Carlip, Class. Quant. Grav. {\bf 17}, 4175
(2000).
\bibitem{das2x} S. Das, P. Majumdar and R.K. Bhaduri, Class. Quant. Grav.
{\bf 19}, 2355 (2002).
\bibitem{sol1x} S.N. Solodukhin, Phys. Rev. {\bf D51},
609 (1995); {\it ibid}, 618 (1995); {\it ibid} {\bf D57}, 2410
(1998); N. E. Mavromatos, E. Winstanley, Phys. Rev. {\bf
D53},3190, (1996).
\bibitem{furx} D.V. Fursaev, Phys. Rev. {\bf D51}, 5352 (1995).
\bibitem{loux} C.O. Lousto, Phys. Lett. {\bf B352}, 228 (1995).
\bibitem{frox} V.P. Frolov, W.Israel and S.N. Solodukhin,
Phys. Rev. {\bf D54}, 2732 (1996).
\bibitem{sol2x} R.B. Mann and S.N. Solodukhin,
Phys. Rev. {\bf D55}, 3622 (1997);  Nucl. Phys. {\bf B523},
293(1998).
\bibitem{kasx} H.A. Kastrup,  Phys. Lett. {\bf B413}, 267 (1997).
\bibitem{kun1x} A.J.M.  Medved and G. Kunstatter, Phys. Rev. {\bf D60},
104029 (1999).
\bibitem{gg1x} G. Gour, Phys. Rev. {\bf D61}, 021501 (2000).
\bibitem{ostx} O. Obregon, M. Sabido and V.I. Tkach, Gen. Rel. Grav.
{\bf 33},  913 (2001).
\bibitem{jinx}
 J. Jing and M.-L. Yan, Phys. Rev. {\bf D63}, 024003 (2001).
\bibitem{das1x}
 S. Das, R.K. Kaul and P. Majumdar, Phys. Rev. {\bf D63}, 044019 (2001).
\bibitem{bir1x}
D.Birmingham and S. Sen, Phys. Rev. {\bf D63}, 047501 (2001).
\bibitem{maj2x} P. Majumdar, Pramana {\bf 55}, 511 (2000); hep-th/0110198 (2001).
\bibitem{kun2x} A.J.M. Medved and G. Kunstatter, Phys. Rev. {\bf D63},
104005 (2001).
\bibitem{ms2x} S.A. Major and K.L. Setter, Class. Quant. Grav.
{\bf 18}, 5125 (2001); {\it ibid} 5293 (2001).
\bibitem{bir2x} D. Birmingham, I. Sachs and S. Sen,
Int. J. Mod. Phys. {\bf D10}, 833 (2001).
\bibitem{govx}
T.R. Govindarajan, R.K. Kaul and V. Suneeta, Class. Quant. Grav.
{\bf 18}, 2877 (2001).
\bibitem{cavx} M. Cavaglia and A. Fabbri, Phys. Rev. {\bf D65},
044012 (2002).
\bibitem{gup1x} K.S. Gupta and S. Sen, Phys. Lett. {\bf B526}, 121 (2002).
\bibitem{ajm1x} A.J.M. Medved, Class. Quant. Grav. {\bf 19}, 2503
(2002).
\bibitem{gup2x} K.S. Gupta, hep-th/0204137.
\bibitem{mukx} S. Mukherji and S.S. Pal, JHEP {\bf 0205}, 026
(2002).
\bibitem{das3x} S. Das, hep-th/0207072.
\bibitem{ajm2x} A.J.M. Medved, hep-th/0210017, gr-qc/0211004.
\bibitem{gg2x} G. Gour, Phys. Rev. {\bf D66}, 104022 (2002).
\bibitem{kripx} I.B. Khriplovich, gr-qc/0210108.
\bibitem{kaul1x} R.K. Kaul and S.K. Rama, gr-qc/0301128.
\bibitem{kaul2x} R.K. Kaul, hep-th/0302170.
\bibitem{maj3x} A. Chatterjee and P. Majumdar, gr-qc/0303030.
\bibitem{das4x} S. Das and V. Husain, hep-th/0303089.
\bibitem{das5x} M.M. Akbar and S. Das, hep-th/0304076.
\bibitem{od}J. E. Lidsey, S. Nojiri, S. D. Odintsov and S. Ogushi,
Phys. Lett. {\bf B544}, 337, (2002).
\bibitem{medved}G. Gour, A. J. M. Medved, gr-qc/0305018.
\bibitem{od1} S. Nojiri, S. D. Odintsov, S. Ogushi,
hep-th/0212047.

\end{thebibliography}
\end{document}